\documentclass[twocolumn,aps,tightenlinefs,superscriptaddress,preprintnumbers]{revtex4}
\usepackage{graphicx,color,dcolumn,booktabs,bm}
\usepackage{longtable,lscape}
\usepackage{amsmath}
\usepackage{indentfirst}
\usepackage{epsfig}
\usepackage{feynmf}   
\usepackage{epstopdf}   
\usepackage{slashed}  
\usepackage{cases}
\usepackage{color}
\usepackage{multirow}
\usepackage{ulem}
\usepackage{verbatim}
\usepackage[colorlinks,linkcolor=red,anchorcolor=green,citecolor=blue]{hyperref}
\usepackage{mathrsfs}
\usepackage{rotating}
\usepackage{threeparttable}
\usepackage{lineno}
\usepackage{subfigure}

\graphicspath{{ps}}

\newcommand{\ra}{\rightarrow}
\newcommand{\EE}{e^+e^-}
\newcommand{\lamc}{\Lambda_c^+}

\newcommand{\pkpi}{pK^-\pi^+}
\newcommand{\ppiz}{p\pi^0}
\newcommand{\peta}{p\eta}
\newcommand{\GG}{\gamma\gamma}
\newcommand{\BF}{\mathcal{B}}

\begin{document}



\title{ \quad\\[0.1cm] \boldmath Measurements of the branching fractions of $\Lambda_c^+ \to p \eta$ and $\Lambda_c^+ \to p \pi^0$ decays at Belle}

\noaffiliation
\affiliation{Department of Physics, University of the Basque Country UPV/EHU, 48080 Bilbao}
\affiliation{University of Bonn, 53115 Bonn}
\affiliation{Brookhaven National Laboratory, Upton, New York 11973}
\affiliation{Budker Institute of Nuclear Physics SB RAS, Novosibirsk 630090}
\affiliation{Faculty of Mathematics and Physics, Charles University, 121 16 Prague}
\affiliation{Chonnam National University, Gwangju 61186}
\affiliation{University of Cincinnati, Cincinnati, Ohio 45221}
\affiliation{Deutsches Elektronen--Synchrotron, 22607 Hamburg}
\affiliation{Duke University, Durham, North Carolina 27708}
\affiliation{University of Florida, Gainesville, Florida 32611}
\affiliation{Department of Physics, Fu Jen Catholic University, Taipei 24205}
\affiliation{Key Laboratory of Nuclear Physics and Ion-beam Application (MOE) and Institute of Modern Physics, Fudan University, Shanghai 200443}
\affiliation{Justus-Liebig-Universit\"at Gie\ss{}en, 35392 Gie\ss{}en}
\affiliation{Gifu University, Gifu 501-1193}
\affiliation{SOKENDAI (The Graduate University for Advanced Studies), Hayama 240-0193}
\affiliation{Gyeongsang National University, Jinju 52828}
\affiliation{Department of Physics and Institute of Natural Sciences, Hanyang University, Seoul 04763}
\affiliation{University of Hawaii, Honolulu, Hawaii 96822}
\affiliation{High Energy Accelerator Research Organization (KEK), Tsukuba 305-0801}
\affiliation{J-PARC Branch, KEK Theory Center, High Energy Accelerator Research Organization (KEK), Tsukuba 305-0801}
\affiliation{Higher School of Economics (HSE), Moscow 101000}
\affiliation{Forschungszentrum J\"{u}lich, 52425 J\"{u}lich}
\affiliation{IKERBASQUE, Basque Foundation for Science, 48013 Bilbao}
\affiliation{Indian Institute of Science Education and Research Mohali, SAS Nagar, 140306}
\affiliation{Indian Institute of Technology Bhubaneswar, Satya Nagar 751007}
\affiliation{Indian Institute of Technology Guwahati, Assam 781039}
\affiliation{Indian Institute of Technology Hyderabad, Telangana 502285}
\affiliation{Indian Institute of Technology Madras, Chennai 600036}
\affiliation{Indiana University, Bloomington, Indiana 47408}
\affiliation{Institute of High Energy Physics, Chinese Academy of Sciences, Beijing 100049}
\affiliation{Institute of High Energy Physics, Vienna 1050}
\affiliation{Institute for High Energy Physics, Protvino 142281}
\affiliation{INFN - Sezione di Napoli, 80126 Napoli}
\affiliation{Advanced Science Research Center, Japan Atomic Energy Agency, Naka 319-1195}
\affiliation{J. Stefan Institute, 1000 Ljubljana}
\affiliation{Institut f\"ur Experimentelle Teilchenphysik, Karlsruher Institut f\"ur Technologie, 76131 Karlsruhe}
\affiliation{Kavli Institute for the Physics and Mathematics of the Universe (WPI), University of Tokyo, Kashiwa 277-8583}
\affiliation{Kitasato University, Sagamihara 252-0373}
\affiliation{Korea Institute of Science and Technology Information, Daejeon 34141}
\affiliation{Korea University, Seoul 02841}
\affiliation{Kyoto Sangyo University, Kyoto 603-8555}
\affiliation{Kyungpook National University, Daegu 41566}
\affiliation{Universit\'{e} Paris-Saclay, CNRS/IN2P3, IJCLab, 91405 Orsay}
\affiliation{P.N. Lebedev Physical Institute of the Russian Academy of Sciences, Moscow 119991}
\affiliation{Liaoning Normal University, Dalian 116029}
\affiliation{Faculty of Mathematics and Physics, University of Ljubljana, 1000 Ljubljana}
\affiliation{Ludwig Maximilians University, 80539 Munich}
\affiliation{Luther College, Decorah, Iowa 52101}
\affiliation{Malaviya National Institute of Technology Jaipur, Jaipur 302017}
\affiliation{University of Maribor, 2000 Maribor}
\affiliation{Max-Planck-Institut f\"ur Physik, 80805 M\"unchen}
\affiliation{School of Physics, University of Melbourne, Victoria 3010}
\affiliation{University of Mississippi, University, Mississippi 38677}
\affiliation{University of Miyazaki, Miyazaki 889-2192}
\affiliation{Moscow Physical Engineering Institute, Moscow 115409}
\affiliation{Graduate School of Science, Nagoya University, Nagoya 464-8602}
\affiliation{Kobayashi-Maskawa Institute, Nagoya University, Nagoya 464-8602}
\affiliation{Universit\`{a} di Napoli Federico II, 80126 Napoli}
\affiliation{Nara Women's University, Nara 630-8506}
\affiliation{National Central University, Chung-li 32054}
\affiliation{National United University, Miao Li 36003}
\affiliation{Department of Physics, National Taiwan University, Taipei 10617}
\affiliation{H. Niewodniczanski Institute of Nuclear Physics, Krakow 31-342}
\affiliation{Nippon Dental University, Niigata 951-8580}
\affiliation{Niigata University, Niigata 950-2181}
\affiliation{University of Nova Gorica, 5000 Nova Gorica}
\affiliation{Novosibirsk State University, Novosibirsk 630090}
\affiliation{Osaka City University, Osaka 558-8585}
\affiliation{Pacific Northwest National Laboratory, Richland, Washington 99352}
\affiliation{Panjab University, Chandigarh 160014}
\affiliation{Peking University, Beijing 100871}
\affiliation{University of Pittsburgh, Pittsburgh, Pennsylvania 15260}
\affiliation{Research Center for Nuclear Physics, Osaka University, Osaka 567-0047}
\affiliation{Meson Science Laboratory, Cluster for Pioneering Research, RIKEN, Saitama 351-0198}
\affiliation{Department of Modern Physics and State Key Laboratory of Particle Detection and Electronics, University of Science and Technology of China, Hefei 230026}
\affiliation{Seoul National University, Seoul 08826}
\affiliation{Showa Pharmaceutical University, Tokyo 194-8543}
\affiliation{Soochow University, Suzhou 215006}
\affiliation{Soongsil University, Seoul 06978}
\affiliation{Sungkyunkwan University, Suwon 16419}
\affiliation{School of Physics, University of Sydney, New South Wales 2006}
\affiliation{Department of Physics, Faculty of Science, University of Tabuk, Tabuk 71451}
\affiliation{Tata Institute of Fundamental Research, Mumbai 400005}
\affiliation{Department of Physics, Technische Universit\"at M\"unchen, 85748 Garching}
\affiliation{Department of Physics, Tohoku University, Sendai 980-8578}
\affiliation{Earthquake Research Institute, University of Tokyo, Tokyo 113-0032}
\affiliation{Department of Physics, University of Tokyo, Tokyo 113-0033}
\affiliation{Tokyo Institute of Technology, Tokyo 152-8550}
\affiliation{Tokyo Metropolitan University, Tokyo 192-0397}
\affiliation{Utkal University, Bhubaneswar 751004}
\affiliation{Virginia Polytechnic Institute and State University, Blacksburg, Virginia 24061}
\affiliation{Wayne State University, Detroit, Michigan 48202}
\affiliation{Yamagata University, Yamagata 990-8560}
\affiliation{Yonsei University, Seoul 03722}
  \author{S.~X.~Li}\affiliation{Key Laboratory of Nuclear Physics and Ion-beam Application (MOE) and Institute of Modern Physics, Fudan University, Shanghai 200443} 
  \author{C.~P.~Shen}\affiliation{Key Laboratory of Nuclear Physics and Ion-beam Application (MOE) and Institute of Modern Physics, Fudan University, Shanghai 200443} 
  \author{I.~Adachi}\affiliation{High Energy Accelerator Research Organization (KEK), Tsukuba 305-0801}\affiliation{SOKENDAI (The Graduate University for Advanced Studies), Hayama 240-0193} 
  \author{J.~K.~Ahn}\affiliation{Korea University, Seoul 02841} 
  \author{H.~Aihara}\affiliation{Department of Physics, University of Tokyo, Tokyo 113-0033} 
  \author{D.~M.~Asner}\affiliation{Brookhaven National Laboratory, Upton, New York 11973} 
  \author{T.~Aushev}\affiliation{Higher School of Economics (HSE), Moscow 101000} 
  \author{R.~Ayad}\affiliation{Department of Physics, Faculty of Science, University of Tabuk, Tabuk 71451} 
  \author{V.~Babu}\affiliation{Deutsches Elektronen--Synchrotron, 22607 Hamburg} 
  \author{S.~Bahinipati}\affiliation{Indian Institute of Technology Bhubaneswar, Satya Nagar 751007} 
% ...
\collaboration{The Belle Collaboration}


\begin{abstract}
We report measurements of the branching fractions of singly Cabibbo-suppressed decays $\Lambda_c^+ \to p \eta$ and $\Lambda_c^+ \to p \pi^0$
using the full Belle data sample corresponding to an integrated luminosity of 980.6 $\rm fb^{-1}$.
The data were collected by the Belle detector at the KEKB $e^{+}$$e^{-}$ asymmetric-energy collider.
A clear $\Lambda_c^+$ signal is seen in the invariant mass distribution of $p \eta$.
The fitted number of signal events of the $\Lambda_c^+ \to p \eta$ process is $7734 \pm 263$; from this, we measure the ratio of branching fractions
${\cal B}(\Lambda_c^+ \to p \eta)/{\cal B}(\Lambda_c^+ \to p K^- \pi^+) =  (2.258  \pm 0. 077(\rm stat. ) \pm  0.122(\rm syst. ))\times 10^{-2}$, from which we infer
the branching fraction ${\cal B}(\Lambda_c^+ \to p \eta) = (1.42 \pm 0.05(\rm stat.) \pm 0.11(\rm syst.)) \times 10^{-3}$.
In addition, no significant signal for $\Lambda_c^+ \to p \pi^0$ is found so
an upper limit on the branching fraction of ${\cal B}(\Lambda_c^+ \to p \pi^0)<8.0 \times 10^{-5}$ at 90\% credibility level is set,
more than three times better than the best current upper limit.
\end{abstract}

\maketitle

\tighten

\section{\boldmath Introduction}

Weak decays of charmed baryons are useful for testing many contradictory theoretical models and methods, e.g.\ the flavor symmetry approach and heavy quark effective theory~\cite{lamcR1,lamcR2, lamcR3, lamcR4}.
In contrast with the decays of charmed mesons, the decays of some charmed baryons are helicity suppressed, making the $W$-boson exchange favored~\cite{lamcR5}.
The understanding of charmed baryons has progressed relatively slowly compared to that of charmed mesons.
The main reason is that the cross section for the generation of charmed baryons is smaller than that of the mesons, so that
some reactions with small decay branching fractions are difficult to observe experimentally~\cite{lamcR6,lamcR7,lamcR8}.
Although there have been many improved measurements of the properties of charmed baryons, precision measurements of the decay branching fractions still remain poor for many Cabibbo favored (CF) decay modes and even worse for some decay modes dominated by Cabibbo suppression and $W$-boson exchange~\cite{pdg}.

\begin{figure}[h]
             \centering
             \vspace{-0.2cm}
             \subfigtopskip=1pt
             \subfigbottomskip=1pt
             \subfigcapskip=-5pt
             \subfigure[]{
                          \begin{minipage}[t]{0.5\linewidth}
                          \centering
                          \includegraphics[width=0.9\textwidth]{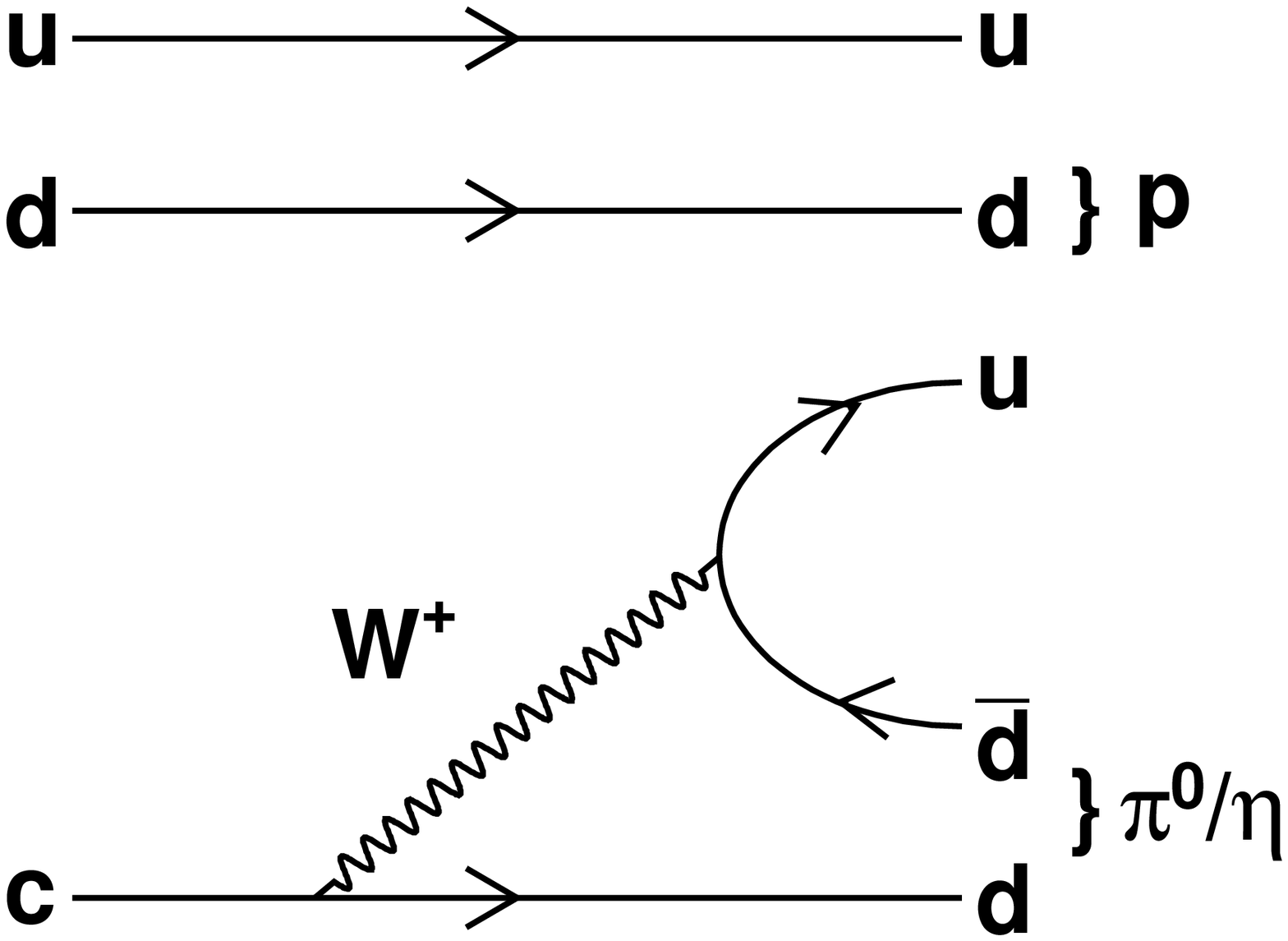}
                          \end{minipage}%
             }%
             \subfigure[]{
                          \begin{minipage}[t]{0.5\linewidth}
                          \flushleft
                          \includegraphics[width=0.9\textwidth]{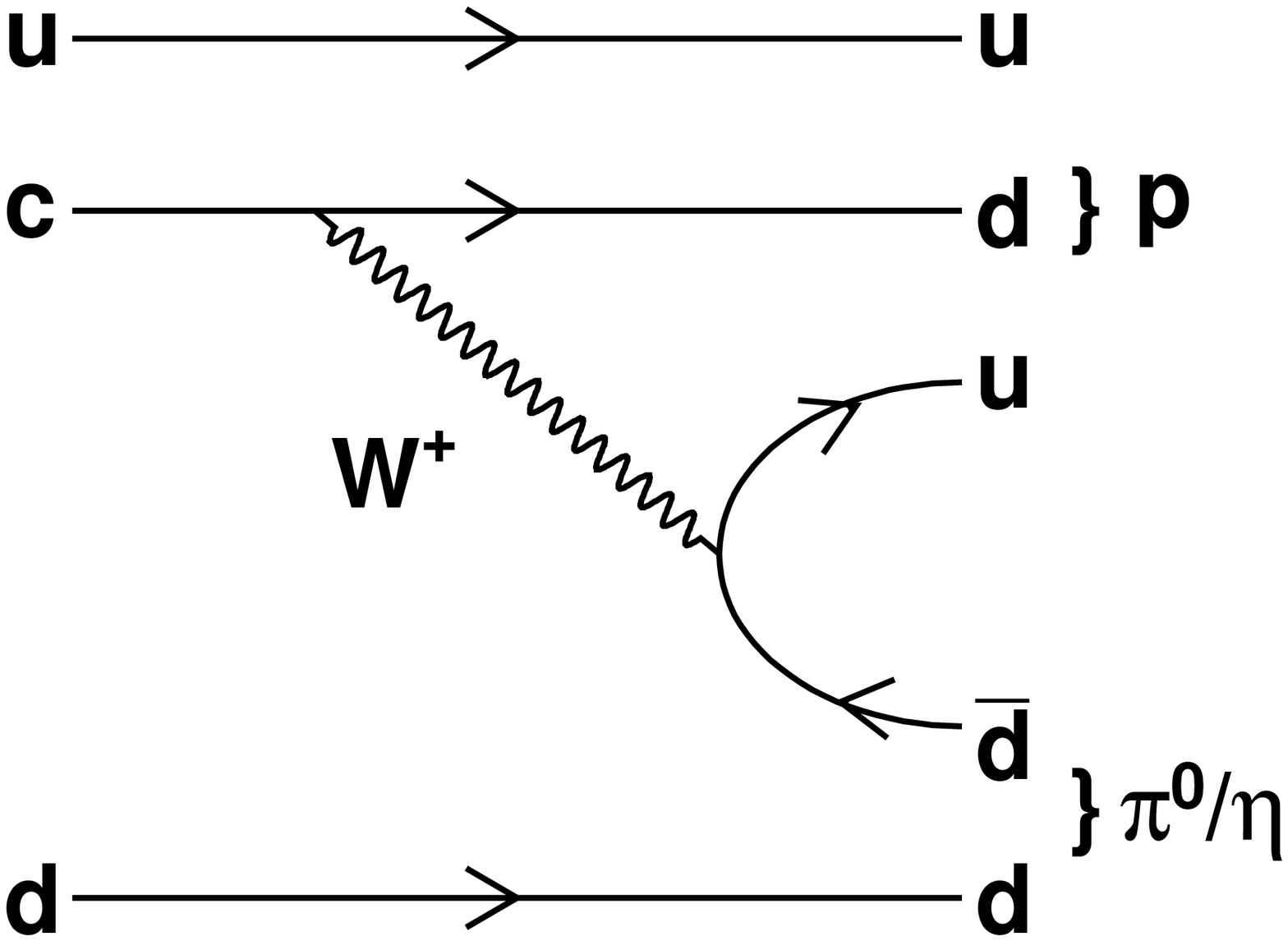}
                          \end{minipage}%
             }%
             \quad
             \subfigure[]{
                          \begin{minipage}[t]{0.5\linewidth}
                          \centering
                          \includegraphics[width=0.9\textwidth]{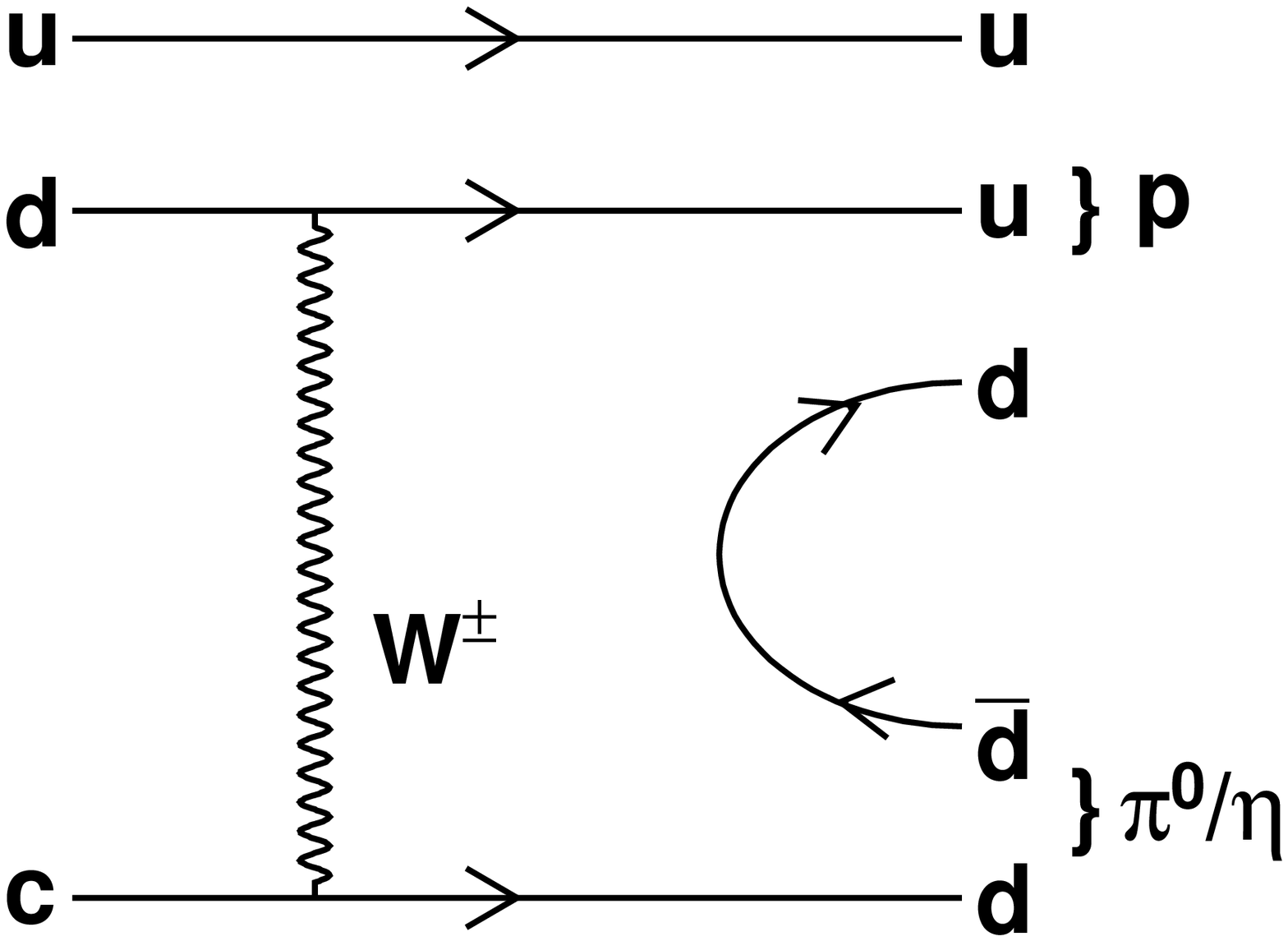}
                          \end{minipage}%
             }%
             \subfigure[]{
                          \begin{minipage}[t]{0.5\linewidth}
                          \centering
                          \includegraphics[width=0.9\textwidth]{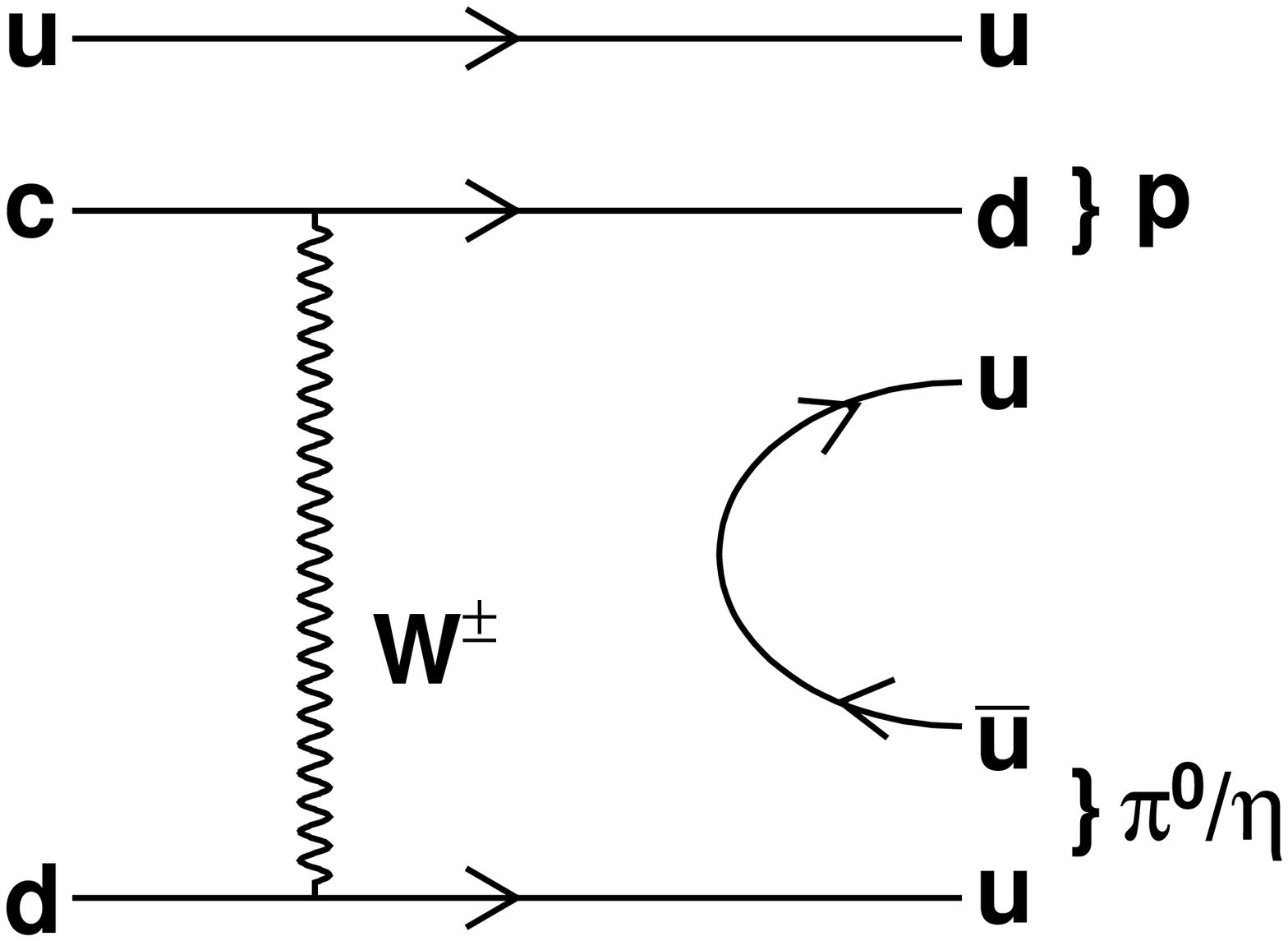}
                          \end{minipage}%
             }%
             \quad
             \subfigure[]{
                          \begin{minipage}[t]{0.5\linewidth}
                          \centering
                          \includegraphics[width=0.9\textwidth]{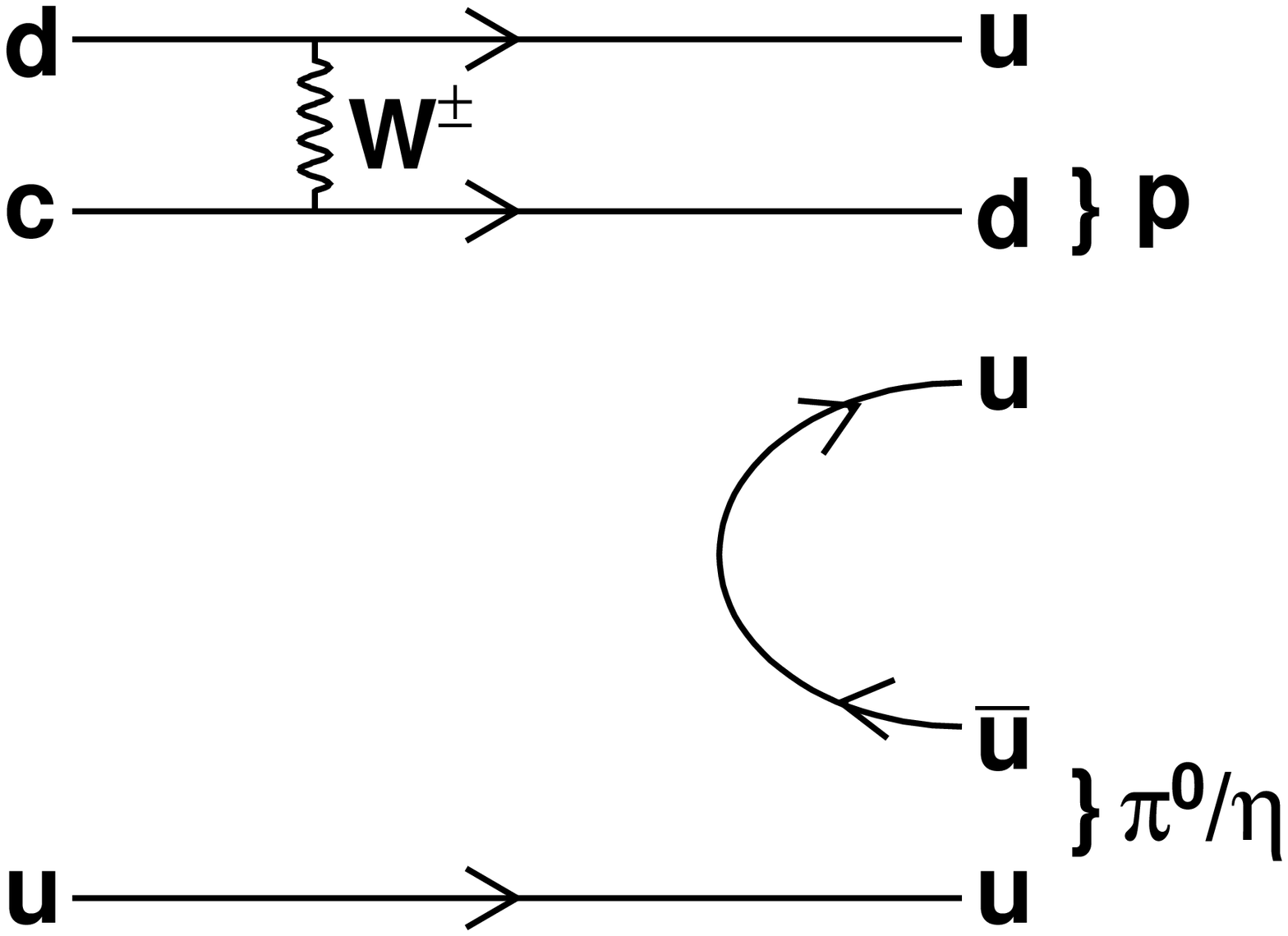}
                          \end{minipage}%
             }%
             \subfigure[]{
                          \begin{minipage}[t]{0.5\linewidth}
                          \centering
                          \includegraphics[width=0.9\textwidth]{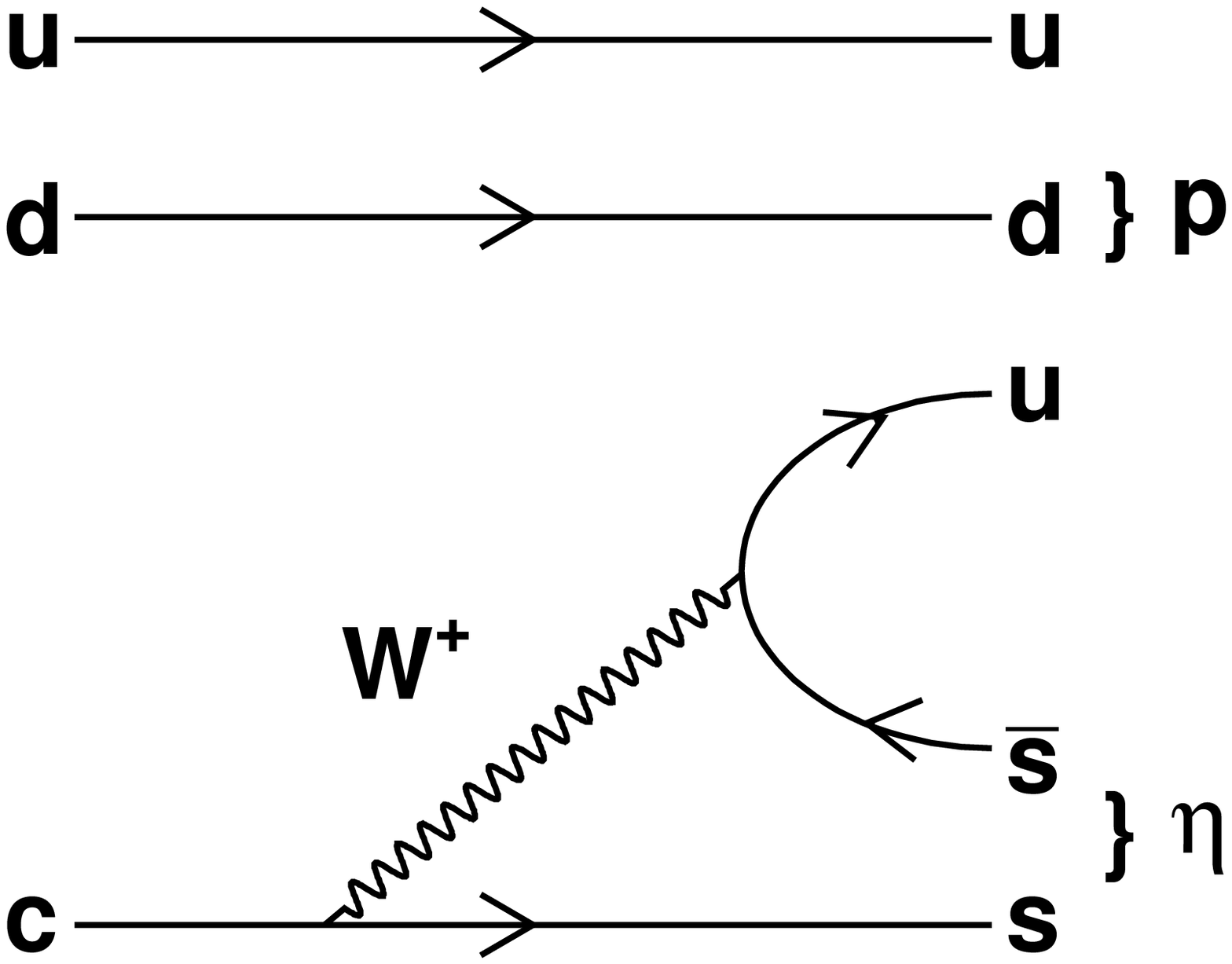}
                          \end{minipage}%
             }%
             \centering
             \caption{Feynman diagrams for the weak Cabibbo-suppressed decays $\Lambda_c^+ \to p \pi^0$ and $\Lambda_c^+ \to p \eta$:
             (a,~b) Internal $W$ emission,  (c,~d,~e) $W$ exchange, and (f) Internal $W$ emission for $\lamc \ra \peta$.}
             \label{feynman}
      \end{figure}

In theory, the singly Cabibbo-suppressed (SCS) decays $\lamc \ra \ppiz$ and $\lamc \ra \peta$ proceed predominantly through internal
$W$ emission and $W$ exchange.
Typical decay diagrams of two SCS decays are shown in Fig.~\ref{feynman}.
The internal $W$ emission involving an $s$ quark in Fig.~\ref{feynman}(f) is allowed in $\lamc \ra \peta$ but absent in $\lamc \ra \ppiz$. The theoretical calculations predict the branching fraction of $\lamc \ra \peta$ at least an order of magnitude greater than that of $\lamc \ra \ppiz$
and give different assumption-dependent results for the branching fractions of these SCS decays~\cite{lamcR1,lamcR3,lamcR9,lamcR10,lamcR11}.
In contrast with the strong decays of heavy-flavor mesons, the $W$-boson exchange mechanism plays an important role in the decay of charmed baryons.
Thus, measuring the branching fractions of these two SCS decays will help elucidate the decay mechanism of charmed baryons.

The first evidence for the decay $\lamc \ra \peta$ with a statistical significance of 4.2$\sigma$
and a branching fraction of $\BF(\lamc \ra \peta) = (1.24\pm0.30)\times 10^{-3}$ was reported by the BESIII Collaboration~\cite{besResult}.
They found no significant $\lamc \ra \ppiz$ signal and
set an upper limit on its branching fraction $\BF(\lamc \ra \ppiz) < 2.7\times 10^{-4}$ at 90\% confidence level~\cite{besResult}.

To improve the measurement precision, we measure the ratio of the branching fractions of the two SCS processes  with respect to the CF $\lamc \ra \pkpi$ decay mode:
\begin{small}
\begin{equation}
\label{bf-cal}
\frac{\BF(\rm SCS)}{\BF(\rm CF)}=\frac{N^{\rm obs}(\rm SCS)}{\epsilon^{\rm MC}(\rm SCS)\times\BF(\pi^{0}/\eta \ra \GG)}\times\frac{\epsilon^{MC}(\rm CF)}{N^{\rm obs}(\rm CF)},
\end{equation}
\end{small}
where $\BF$, $\epsilon^{\rm MC}$, and $N^{\rm obs}$ are the branching fraction, signal efficiency, and the fitted yield of signal events from data, respectively.
The value of the branching fraction of the CF decay is $(6.28 \pm 0.32) \times 10^{-2}$~\cite{pdg}.
The values of $\BF(\pi^{0} \ra \GG)$ and $\BF(\eta \ra \GG)$ are $0.9882\pm0.0003$ and $0.3941\pm0.0020$, respectively~\cite{pdg}.

\section{\boldmath The data sample and the belle detector}

The measurements of the two SCS branching fractions are based on a data sample taken at or near
the $\Upsilon(1S)$, $\Upsilon(2S)$, $\Upsilon(3S)$, $\Upsilon(4S)$, and $\Upsilon(5S)$ resonances
collected with the Belle detector at the KEKB asymmetric-energy $e^+e^-$ collider~\cite{KEKB}.
The integrated luminosity of the data samples is 980.6 $\rm fb^{-1}$,
including 711 $\rm fb^{-1}$ on the $\Upsilon(4S)$ resonance,
89.4 $\rm fb^{-1}$ off the $\Upsilon(4S)$ resonance, 121.4 $\rm fb^{-1}$ on the $\Upsilon(5S)$ resonance,
and 58.8 $\rm fb^{-1}$ at the $\Upsilon(1S,~2S,~3S)$ resonances.
The Belle detector is a large-solid-angle magnetic spectrometer that
consists of a silicon vertex detector (SVD), a 50-layer central drift chamber (CDC),
an array of aerogel threshold Cherenkov counters (ACC),
a barrel-like arrangement of time-of-flight scintillation counters (TOF),
and an electromagnetic calorimeter comprised of CsI(Tl) crystals (ECL) located inside
a superconducting solenoid coil that provides a 1.5~T magnetic field.
An iron flux-return located outside of the coil is instrumented to detect $K_L^0$ mesons and to identify muons (KLM).  The detector is described in detail elsewhere~\cite{Belle}.

Signal MC samples of $\EE \ra c\bar{c}$; $c\bar{c} \ra \lamc X$ with $X$ denoting anything; $\lamc \ra \pkpi/\ppiz/\peta$ are
used to optimize the selection criteria and estimate the reconstruction and selection efficiency,
and are generated under the $\Upsilon(4S)$ resonance condition with {\sc pythia}~\cite{pythia} and {\sc EvtGen}~\cite{evtgen} and propagated with {\sc geant3}~\cite{geant3} to simulate the detector performance.
The charged-conjugate modes are included unless otherwise stated.

Inclusive MC samples of $\Upsilon(4S)\ra B^{+}B^{-}/B^{0}\bar{B}^{0}$, $\Upsilon(5S)\ra B_{s}^{(*)}\bar{B}_{s}^{(*)}$, $\EE \to q\bar{q}$ $(q=u,~d,~s,~c)$ at $\sqrt{s}$ = 10.58 and 10.867~GeV, and $\Upsilon(1S,~2S,~3S)$ decays corresponding to two times the integrated luminosity of each data set are used to
characterize the (potentially peaking) backgrounds~\cite{topoana}.

\section{\boldmath Event selection criteria}

For charged-particle tracks, the distance of closest approach with respect to the interaction point (IP)
along the $z$ axis (parallel to the positron beam) and in the transverse $r\phi$ plane is required to be less than 2.0 cm and 0.1 cm, respectively.
In addition, each track is required to have at least one SVD hit.
Particle identification (PID) is used to discriminate the type of charged hadron tracks:
$\mathcal{R}(h|h') = \mathcal{L}(h)/(\mathcal{L}(h)+\mathcal{L}(h'))$ is defined as the ratio of the likelihoods for the $h$ and $h'$
hypotheses, where $\mathcal{L}(h)$ ($h$ = $\pi$, $K$, or $p$) is the combined likelihood derived from the ACC, TOF, and CDC $dE/dx$ measurements~\cite{pidcode}.
$\mathcal{R}(p|\pi)>0.9$ and $\mathcal{R}(p|K)>0.9$ are required for protons.
$\mathcal{R}(K|p) > 0.4$ and $\mathcal{R}(K|\pi) > 0.9$ are required for charged kaons.
$\mathcal{R}(\pi|p) > 0.4$ and $\mathcal{R}(\pi|K) > 0.4$ are required for charged pions.
$\mathcal{R}(e)$, a likelihood ratio for $e$ and $h$ identification formed from ACC, CDC, and ECL information~\cite{eidcode}, is required to be less than 0.9 for all charged tracks to remove electrons. For the typical momentum range of our SCS decays, the identification efficiencies of $p$, $K$, and $\pi$ are 81.7\%, 79.6\%, and 96.9\%, respectively.

A $\lamc$ candidate for the CF decay is reconstructed from three tracks identified as $p$, $K$, and $\pi$,
subject to a common-vertex fit.
The $\chi^2$ of the vertex fit is required to be less than 40 to reject background from incorrect combinations.
The scaled momentum of the $\lamc$, defined as $x_{p} = {p^{*}}/{\sqrt{E^{2}_{\rm cm}/4 - M^{2}}}$~\cite{speedoflight}, is required to be greater than 0.53 for all $\lamc$ candidates to suppress the combinatorial background, especially from $B$-meson decays. Here, $E_{\rm cm}$ is the center-of-mass (CM) energy, while $p^{*}$ and $M$ are the momentum and invariant mass, respectively, of the $\Lambda_{c}^{+}$ candidates in the CM frame. All of these optimized  selection criteria are taken from Ref.~\cite{dcspkpi}.

An ECL cluster not matching any track is identified as a photon candidate.
Each photon candidate is required to have
a ratio of energy deposited in the central $3\times3$ square of ECL cells to that deposited in the
enclosing $5\times5$ square of cells of $E9/E25 > 0.8$ to reject neutral hadrons.
An optimized figure-of-merit (FOM) study determines that
the energy of photon candidates must exceed 50~MeV and 110~MeV in the barrel and endcap regions of the ECL, respectively, for both photons from $\pi^0 \ra \GG$.
For the $\eta \ra \gamma_1 \gamma_2$ decay, the
$\gamma_1$ ($\gamma_2$) energy must exceed 220 (260)~MeV, 480 (340)~MeV, and 260 (360)~MeV  in the barrel, forward, and backward endcaps, respectively. Two photon candidates are combined to form a $\pi^0/\eta$ candidate and a mass-constrained fit is performed for this candidate. The $\chi^2$ value of
the mass-constrained fit must be less than 7.5 and 4 for $\pi^0$ and $\eta$ candidates, respectively, to suppress the background in which the two-photon invariant mass is far from $\pi^0$ and $\eta$ nominal masses~\cite{pdg}. The momentum in the CM frame
must be greater than 0.69~GeV/$c$ and 0.82~GeV/$c$ for $\pi^0$ and $\eta$ candidates, respectively. All these requirements are optimized.
An SCS $\lamc$ candidate is reconstructed by combining  a proton candidate with a $\pi^0/\eta$ candidate. Again, $x_{p}$ for the $\lamc \ra \ppiz/\peta$ candidates
is required to exceed 0.53.
After applying all the selection criteria, about 0.8\%, 1.4\%, and 1.7\% of the events in the signal region have multiple $\lamc$ candidates for the $\pkpi$, $\peta$, and $\ppiz$ decays, respectively.

The SCS signal region in data is optimized
with the control sample of $\lamc \ra \pkpi$ as well as the $\lamc$ mass sidebands
to the hidden SCS signal region (i.e. the signal region is blinded) by optimizing the ratio $S/\sqrt{S+B}$,
where $S$ and $B$ are the expected number of signal events for SCS decays in data and the number of background events normalized to the signal region, respectively.
$S$ is obtained via
\begin{equation}
\begin{split}
S &= \epsilon^{\rm MC}(\lamc \ra \ppiz/\peta) \times \frac{N^{\rm obs}(\lamc \ra \pkpi)}{\epsilon^{\rm MC}(\lamc \ra \pkpi)}\\
 &\times \frac{\BF(\lamc \ra \ppiz/\peta)\times \BF(\pi^0/\eta \ra \GG)}{\BF(\lamc \ra \pkpi)},
\end{split}
\end{equation}
where $N^{\rm obs}$ and $\epsilon^{\rm MC}$ are the fitted $\lamc$ events of data and the detection efficiency of the signal MC sample, respectively;  $\BF(\lamc \ra \ppiz/\peta)$ are the branching fractions of $2.7\times 10^{-4}$ and $1.24\times 10^{-3}$ for $\lamc \ra \ppiz$ and $\lamc \ra \peta$, respectively~\cite{besResult}; and $\BF(\lamc \ra \pkpi)$ is the branching fraction of the CF decay~\cite{pdg}.

\section{\boldmath efficiency estimation and fit results}

\begin{figure}[htbp]
             \centering
             \includegraphics[width=0.45\textwidth]{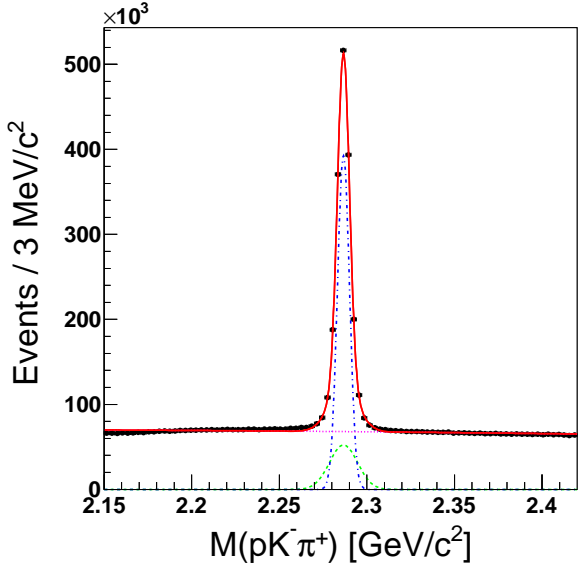}
             \caption{Fit to the invariant mass distribution of $\pkpi$ from data. Black dots with error bars represent the data;  the pink dashed line, the blue dash-dotted line, the green dashed line, and the red solid line represent the background contribution, the core Gaussian, tail Gaussian, and the total fit, respectively.}
             \label{pkpi_data_invM_draw}
  \end{figure}

\begin{figure}[htbp]
             \centering
             \includegraphics[width=0.45\textwidth]{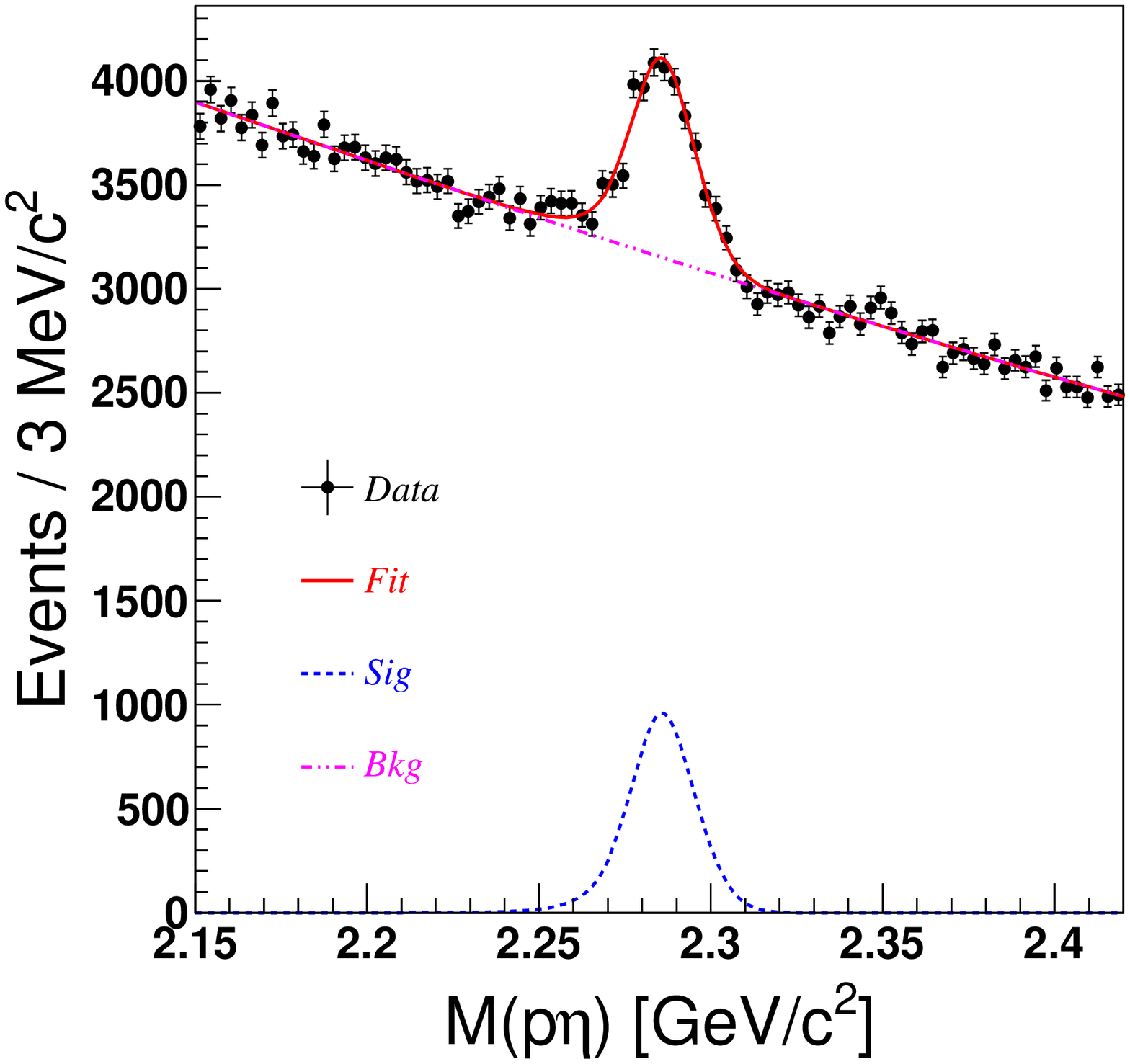}
             \caption{Fit to the invariant mass distribution of $\peta$ from data. Black dots with error bars represent the data; the magenta dash-dotted line, the blue dashed line, and the red solid line represent the background component, the signal, and the total fit, respectively.}
             \label{peta-data-invM-fit}
  \end{figure}

\begin{figure}[htbp]
       \centering
      \includegraphics[width=0.45\textwidth]{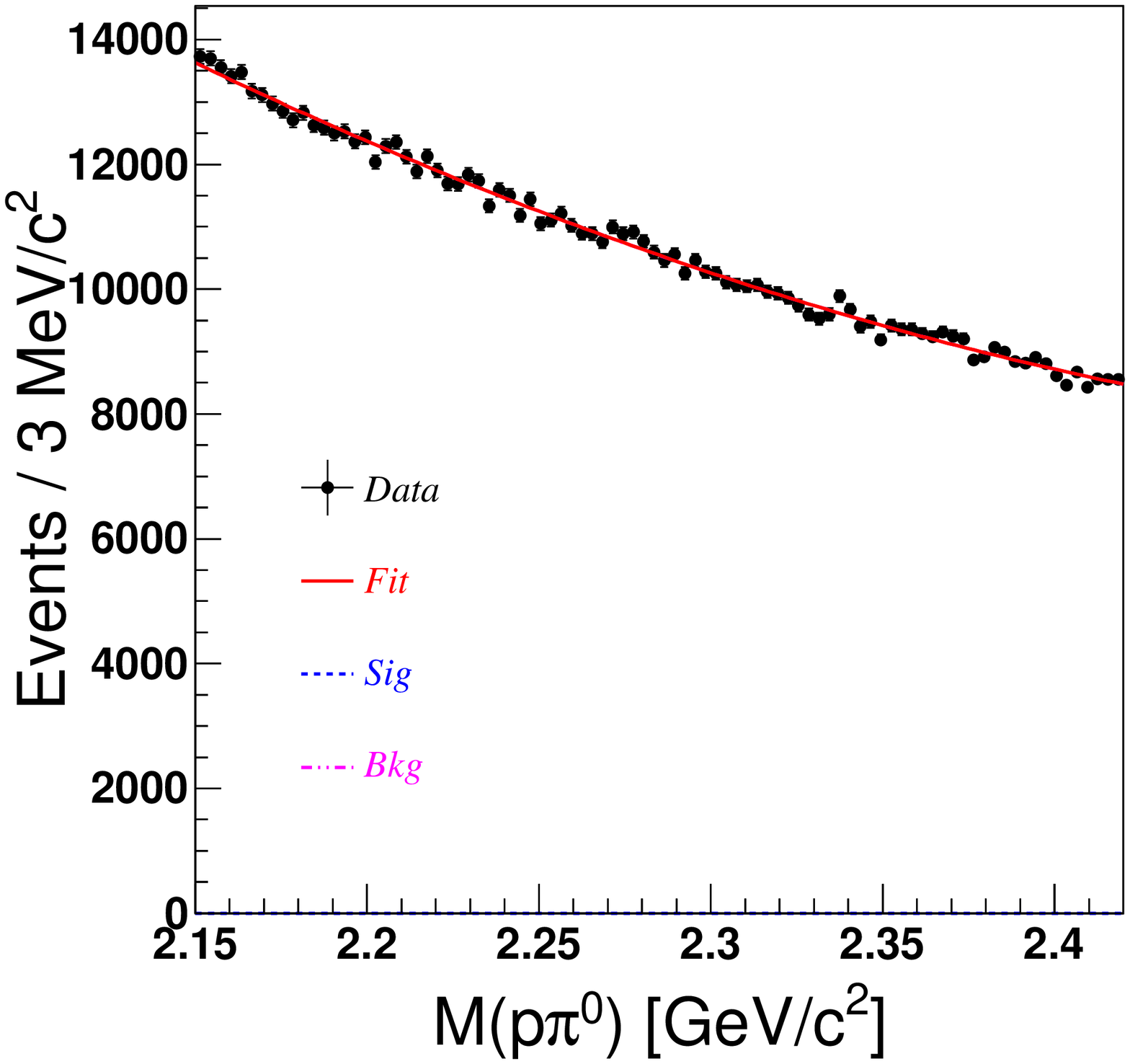}
      \caption{Fit to the invariant mass distribution of $\ppiz$. Black dots with error bars represent the data; the magenta dash-dotted line,
       the blue dashed line, and the red solid line represent the background component, the signal, and the total fit, respectively.}
      \label{ppi0_data_invM_dist_and_fit}
  \end{figure}

With the final selection criteria applied, the invariant mass distributions of $\pkpi$, $\peta$, and $\ppiz$ from data are shown in Figs.~\ref{pkpi_data_invM_draw}, \ref{peta-data-invM-fit}, and \ref{ppi0_data_invM_dist_and_fit}, respectively.
From the study of the topology of inclusive MC samples~\cite{topoana}, no peaking backgrounds contribute to these mass distributions in the $\lamc$ signal region.

For the CF mode, we fit the invariant mass distribution of $\pkpi$ displayed in Fig.~\ref{pkpi_data_invM_draw} from 2.15  to 2.42 GeV/$c^2$ using the binned maximum likelihood fit with a bin width of 3 MeV/$c^{2}$.
A double-Gaussian function with the common mean value is used to model the signal events and a second-order polynomial  is used to model the background events.
The parameters of the signal and background shapes are free in the fit.
The reduced $\chi^{2}$ value of the fit is $\chi^2/{\rm ndf}=87/82=1.06$ and the fitted number of signal events is 1476200 $\pm$ 1560, where $\rm ndf$ is the number of degrees of freedom and the uncertainty is statistical only.

\begin{figure}[htbp]
             \centering
             \includegraphics[width=0.5\textwidth]{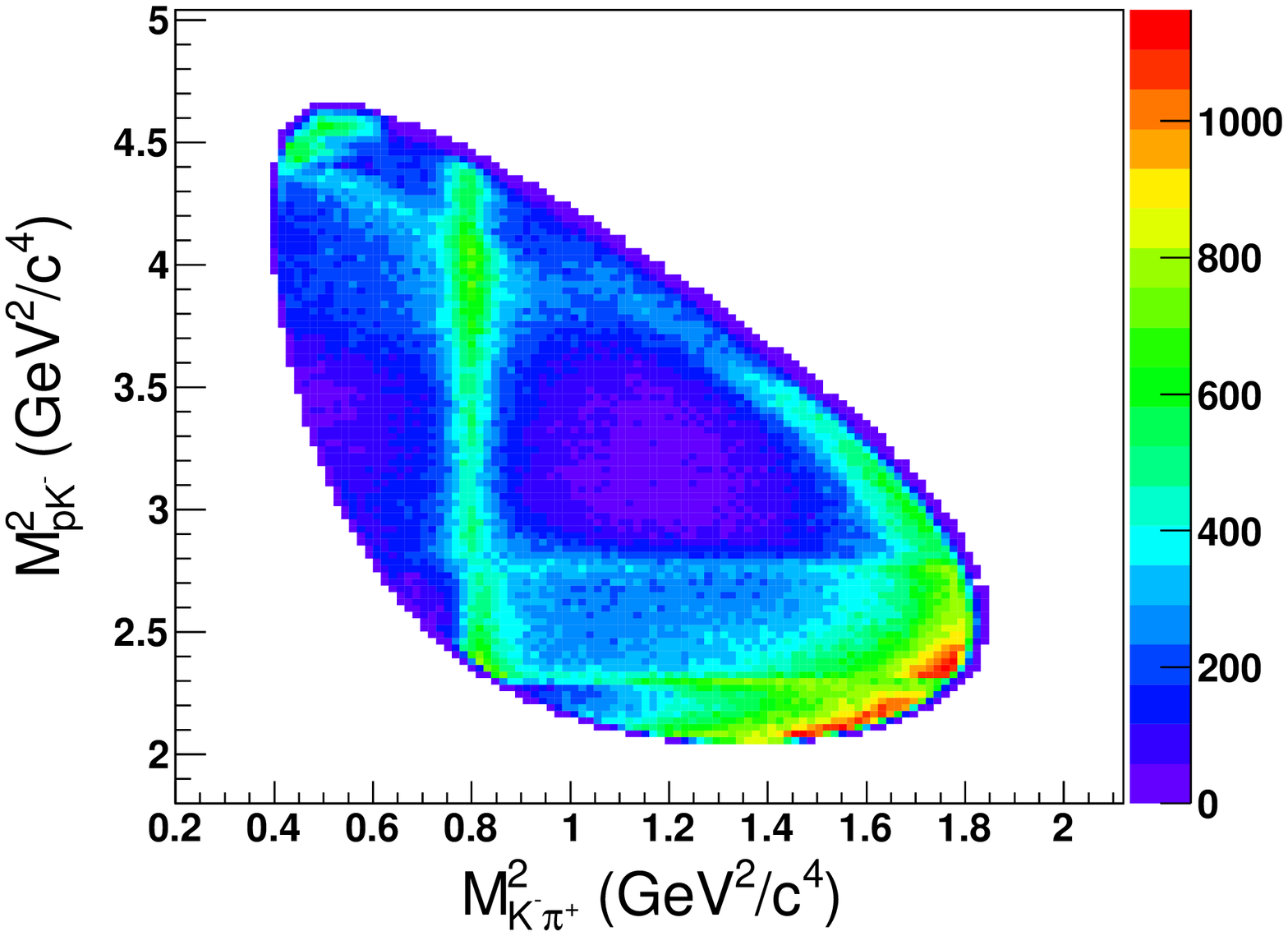}
             \caption{Dalitz plot of the selected $\lamc \ra \pkpi$ candidates.}
             \label{pkpi_data_dalitz}
  \end{figure}

\begin{figure}[htbp]
             \centering
             \includegraphics[width=0.5\textwidth]{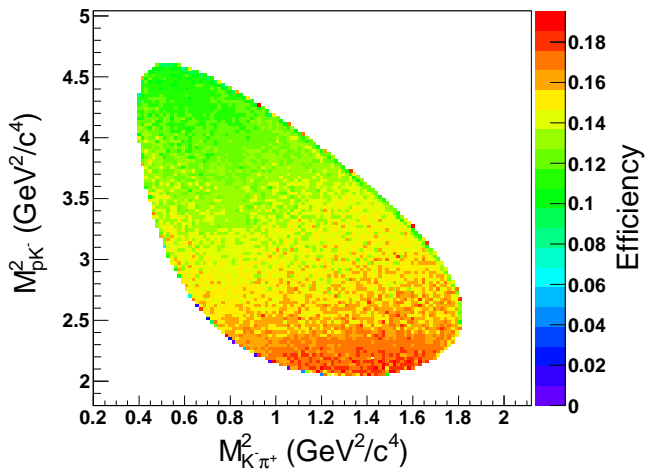}
             \centering
             \caption{Dalitz plot of efficiency distribution for $\lamc \ra \pkpi$ decay.}
             \label{pkpi-sigMC-efficiency-dalitz}
  \end{figure}

The Dalitz~\cite{dalitz} distribution of $M^2(K^-\pi^+)$ versus $M^2(pK^-)$ in the signal region from data is shown in Fig.~\ref{pkpi_data_dalitz}. The signal region is taken from 2.274  to 2.298 GeV/$c^{2}$.
We divide this into 120$\times$120 pixels, with size 0.027 $\rm GeV^{ 2}$$/c^{4}$ for $ M^{2}(pK^{-})$ and 0.016 $\rm GeV^{ 2}$$/c^{4}$ for $M^{2}(K^{-}\pi^{+})$.
The number of background events has been subtracted using the normalized sidebands. The sideband regions are defined to be (2.262, 2.274) GeV/$c^{2}$ and (2.298, 2.310) GeV/$c^{2}$.
A MC sample mixing four subchannels of CF decay weighted with the corresponding branching fractions taken from Ref.~\cite{pdg} is used to assess the selection efficiency of the CF mode. The total number of reconstructed MC signal events is normalized to that of signal candidates in data.
We calculate the overall  efficiency using the efficiency of each pixel.
The formula is $\epsilon = {\Sigma_{i}s_{i}}/{\Sigma_{j}(s_{j}/\epsilon_{j})}$,
where $\Sigma_{i}s_{i}$ is the number of signal candidates in data,
$s_{j}$ and $\epsilon_{j}$ are the number of signal events from data and the efficiency from the MC sample for each pixel, respectively.
The efficiency of one pixel is obtained by dividing the number of events remaining in the signal MC sample by the number of generated events.
The weighted efficiency for each bin is exhibited in Fig.~\ref{pkpi-sigMC-efficiency-dalitz} and the corrected efficiency for data is $(14.06\pm 0.01)$\%.

An obvious $\lamc$ signal peaking in the signal region of the M($\peta$) spectrum is observed.
We use the binned maximum likelihood method to fit the invariant mass distribution of $\peta$ from 2.15  to 2.42 GeV/$c^{2}$ with 3 MeV/$c^{2}$ bin width.
A combined Gaussian and Crystal Ball (CB)~\cite{cb} function with
a common mean value models the signal, and a second-order polynomial models the background.
The parameters of the signal and background line shapes are free in the fit.
Figure~\ref{peta-data-invM-fit} exhibits the distribution of invariant mass of $\peta$ and corresponding fit result.
The reduced $\chi^{2}$ of the fit is $\chi^2/{\rm ndf}=102/83=1.23$ and the fitted number of signal events is $7734 \pm 263$.

\begin{figure}[htbp]
      \centering
      \includegraphics[width=0.5\textwidth]{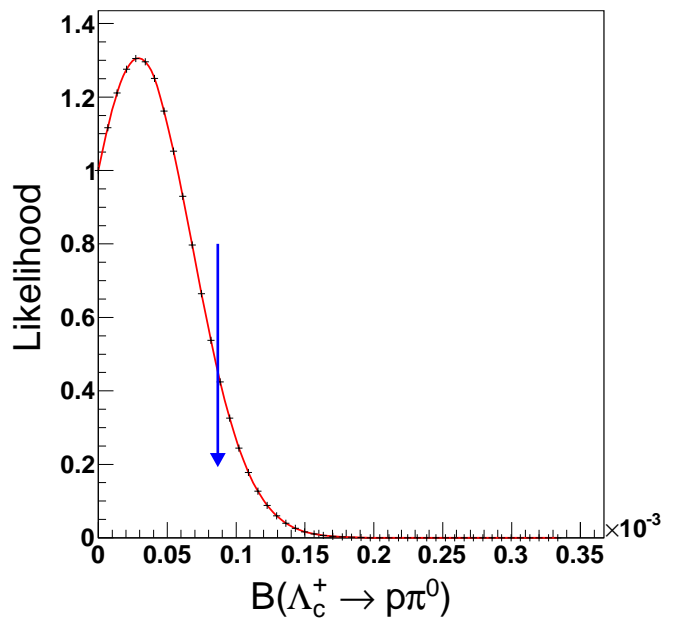}
      \caption{The likelihood distribution as a function of the branching fraction for $\Lambda_c^+ \to p \pi^0$  with the systematic uncertainty included. The blue arrow refers to the 90\% C.L. upper limit on the branching fraction.}
      \label{ppi0-brs-upper-limit}
\end{figure}

There is no significant excess observed in the signal region for $\lamc \ra \ppiz$.
We fit the $M(\ppiz)$ with the binned maximum likelihood method; the fit result is
shown in Fig.~\ref{ppi0_data_invM_dist_and_fit}.
The signal  is modeled by a combined Gaussian and CB function with the
a common mean  convolved with  a Gaussian function; the background  is described by a second-order polynomial.
The parameters of the signal are fixed to MC-derived values and the convolving Gaussian with width 2.1 MeV
accounts for the difference in widths between data and MC
for the  $\lamc \ra \peta$ signal. The fitted number of signal events and the parameters of the background
polynomial are free in the fit.
The fitting range is from 2.15  to 2.42 GeV/$c^{2}$ with a bin width of 3 MeV/$c^{2}$.
The fitted number of signal events is $11\pm 140$, which is consistent with zero.
Thus,  with a uniform prior probability
density function estimation of a Bayesian upper limit is performed to obtain the 90\% credibility level (C.L.) upper limit on the branching fraction of $\lamc \ra \ppiz$.
The likelihood function is integrated from zero to the value that gives 90\% of the total area.
Before integrating, we include the systematic uncertainty ($\sigma_{\rm sys}$) described below by convolving the likelihood distribution with a Gaussian whose width is equal to $\sigma_{\rm sys}$.
An upper limit on the branching fraction of $9.44 \times 10^{-5}$ at 90\% C.L. is set.
The likelihood distribution as a function of the branching fraction, with the systematic uncertainty included, is displayed in Fig.~\ref{ppi0-brs-upper-limit}.

To estimate the efficiencies of the two SCS decays,
we take the ratio of the number of fitted signal events in the invariant mass distribution of $\ppiz/\peta$ to that of generated events from signal MC samples as the efficiency.
We find $(8.28 \pm 0.03)$\% and $(8.89 \pm 0.03)$\% for $\lamc \ra \peta$ and $\lamc \ra \ppiz$, respectively.
The uncertainties are statistical only.

\section{\boldmath Systematic Uncertainties }

Since the branching fraction is obtained from the ratio of the corresponding quantities in Eq.~\ref{bf-cal}, some systematic uncertainties for $\lamc \ra \ppiz/\peta$  cancel. The sources of systematic uncertainties include the fits of CF and SCS decays, PID, tracking efficiency, photon efficiency, the uncertainties of branching fractions of CF and $\pi^{0}/\eta \ra \GG$ decays, and the statistics of the signal MC samples.

To estimate the uncertainties from the fits of CF and SCS decays, we modify the signal and background functions, bin width, and the fit range and refit. To evaluate the uncertainty from the signal function, the signal shape for $\Lambda_c^+ \to p K^- \pi^+$/$p \eta$
is fixed to that from the fit to the MC sample, while that for $\lamc \ra \ppiz$ is changed from a Gaussian and CB combined function to a double CB function. The uncertainty from the background line shape is assessed by using a  first-order polynomial. Furthermore, we change the bin width to 2 MeV/$c^2$ or 4 MeV/$c^2$, and adjust the fit range of invariant mass spectrum to estimate the uncertainties from binning and fit range. The difference of branching fractions between the refitted and
nominal conditions is taken as the uncertainty, which is 3.86\% for $\lamc \ra \ppiz$ and 2.85\% for $\lamc \ra \peta$, respectively.

The systematic uncertainties from PID and tracking efficiency of the proton cancel in the branching-fraction ratio.
Systematic uncertainties of 1.6\% and 1.2\% are assigned for the $K$ and $\pi$ identification efficiencies by studying a low-background control sample of $D^{\ast}$, respectively.
The total systematic uncertainty from PID is 2.0\%, the sum in quadrature of the individual uncertainties for $K$ and $\pi$. From the study of the mid-to-high-momentum track reconstruction efficiency in $D^{*} \ra \pi D^{0}$ decay, the uncertainty of the efficiency for each charged track is 0.35\%, resulting in a total uncertainty of 0.7\% from tracking efficiency. We assign a 2\% systematic uncertainty due to the photon efficiency per photon according to a study of radiative Bhabha events; the total systematic uncertainty from photon reconstruction is thus 4\%.

The systematic uncertainties from the branching fractions of CF and $\pi^{0}/\eta \rightarrow \GG$ are 5.1\%, 0.034\%, and 0.5\%~\cite{pdg}, respectively.

The systematic uncertainty from the size of the signal MC sample is estimated to be 0.34\% and 0.35\% for $\Lambda_{c}^{+} \to p\pi^{0}$ and $\Lambda_{c}^{+} \to p\eta$ decays, respectively.

The systematic uncertainties are summarized in Table \ref{uncertainty-summary}
and give in total 7.8\% and 7.4\% for  $\lamc \ra \ppiz$ and $\lamc \ra \peta$, respectively, which are obtained by assuming all uncertainties are independent and
therefore added  in quadrature.
     \begin{table}[h]
             \centering
             \caption{The sources of the relative systematic uncertainties (\%) on the branching fractions of $\lamc \ra \ppiz$ and $\lamc \ra \peta$ decays.}
             \label{uncertainty-summary}
             \linespread{1.2}
             \begin{tabular}{ c  c  c }   \hline \hline
                         Source  & $\lamc \ra \ppiz$  & $\lamc \ra \peta$ \\ \hline
                         Fit of signal decay                                           &3.9 &2.8    \\
                         PID                                                           &2.0	&2.0    \\
                         Tracking efficiency                                           &0.7	&0.7    \\
                         Photon efficiency                                             &4.0	&4.0    \\
                         $\BF(\lamc \ra \pkpi)$                                        &5.1 &5.1    \\
                         $\BF(\pi^{0}/\eta \ra \GG)$	                               &0.0 &0.5    \\
                         Statistics of signal MC samples                               &0.3 &0.4    \\
                         Total	                                                       &7.8 &7.4    \\ \hline \hline
             \end{tabular}
     \end{table}

\section{\boldmath Conclusion}

We observe the decay $\lamc \ra \peta$.
A significant $\lamc$ signal is observed in the invariant mass distribution of $\peta$ from data.
Using the numbers of the fitted signal events of the $\lamc \ra \peta$ and $\pkpi$ models and the reconstruction efficiencies,
the measured ratio of $\frac{\BF(\lamc \ra \peta)}{\BF(\lamc \ra \pkpi)} = (2.258 \pm 0.077(\rm stat.) \pm 0.122(\rm syst.)) \times 10^{-2}$ is obtained via Eq.(\ref{bf-cal}).
With the independently measured value of $\BF(\lamc \ra \pkpi)$~\cite{pdg}, we extract a branching fraction of $\BF(\lamc \ra \peta) = (1.42 \pm 0.05(\rm stat.) \pm 0.11(\rm syst.)) \times 10^{-3}$, which is consistent with both the latest
published measurement  of $(1.24\pm 0.30 )\times 10^{-3}$~\cite{besResult}, but with much improved precision, and theoretical predictions within 1.3$\sigma$~\cite{lamcR10, lamcR11}.

We see no obvious signal excess in the distribution of $M(\ppiz)$ and so set an
upper limit on the ratio of the branching fractions $\frac{\BF(\lamc \ra \ppiz)}{\BF(\lamc \ra \pkpi)}$ at 90\% C.L.\ of $1.273 \times 10^{-3}$.
From this, we extract  an upper limit on the branching fraction of $\BF(\lamc \ra \ppiz) < 8.0 \times 10^{-5}$ at 90\% C.L., more than three times more stringent than the best current upper
limit of $2.7 \times 10^{-4}$~\cite{besResult}.
The measured $\BF(\lamc \ra \peta)$ is at least an order of magnitude larger than $\BF(\lamc \ra \ppiz)$, which is consistent with the theoretical prediction of internal $W$-emission mechanism involving an $s$ quark in $\lamc \ra \peta$~\cite{lamcR10}.

\section{\boldmath ACKNOWLEDGMENTS}

We thank the KEKB group for the excellent operation of the
accelerator; the KEK cryogenics group for the efficient
operation of the solenoid; and the KEK computer group, and the Pacific Northwest National
Laboratory (PNNL) Environmental Molecular Sciences Laboratory (EMSL)
computing group for strong computing support; and the National
Institute of Informatics, and Science Information NETwork 5 (SINET5) for
valuable network support.  We acknowledge support from
the Ministry of Education, Culture, Sports, Science, and
Technology (MEXT) of Japan, the Japan Society for the
Promotion of Science (JSPS), and the Tau-Lepton Physics
Research Center of Nagoya University;
the Australian Research Council including grants
DP180102629, 
DP170102389, 
DP170102204, 
DP150103061, 
FT130100303; 
Austrian Science Fund (FWF);
the National Natural Science Foundation of China under Contracts
No.~11435013,  
No.~11475187,  
No.~11521505,  
No.~11575017,  
No.~11675166,  
No.~11705209;  
No.~11761141009;
No.~11975076;
No.~12042509;
Key Research Program of Frontier Sciences, Chinese Academy of Sciences (CAS), Grant No.~QYZDJ-SSW-SLH011; 
the  CAS Center for Excellence in Particle Physics (CCEPP); 
the Shanghai Pujiang Program under Grant No.~18PJ1401000;  
the Ministry of Education, Youth and Sports of the Czech
Republic under Contract No.~LTT17020;
the Carl Zeiss Foundation, the Deutsche Forschungsgemeinschaft, the
Excellence Cluster Universe, and the VolkswagenStiftung;
the Department of Science and Technology of India;
the Istituto Nazionale di Fisica Nucleare of Italy;
National Research Foundation (NRF) of Korea Grant
Nos.~2016R1\-D1A1B\-01010135, 2016R1\-D1A1B\-02012900, 2018R1\-A2B\-3003643,
2018R1\-A6A1A\-06024970, 2018R1\-D1A1B\-07047294, 2019K1\-A3A7A\-09033840,
2019R1\-I1A3A\-01058933;
Radiation Science Research Institute, Foreign Large-size Research Facility Application Supporting project, the Global Science Experimental Data Hub Center of the Korea Institute of Science and Technology Information and KREONET/GLORIAD;
the Polish Ministry of Science and Higher Education and
the National Science Center;
the Ministry of Science and Higher Education of the Russian Federation, Agreement 14.W03.31.0026; 
University of Tabuk research grants
S-1440-0321, S-0256-1438, and S-0280-1439 (Saudi Arabia);
the Slovenian Research Agency;
Ikerbasque, Basque Foundation for Science, Spain;
the Swiss National Science Foundation;
the Ministry of Education and the Ministry of Science and Technology of Taiwan;
and the United States Department of Energy and the National Science Foundation.


%


\begin{thebibliography}{99}
\bibitem{speedoflight} We used units in which the speed of light is $c=1$.
\end{thebibliography}
\end{document}